# Electrostatic effects on contacts to carbon nanotube transistors

Aron W. Cummings and François Léonard

Sandia National Laboratories, MS9161, Livermore, California 94551, USA

We use numerical simulations to investigate the effect of electrostatics on the source and drain contacts of carbon nanotube field-effect transistors. We find that unscreened charge on the nanotube at the contact-channel interface leads to a potential barrier that can significantly hamper transport through the device. This effect is largest for intermediate gate voltages and for contacts near the ohmic-Schottky crossover, but can be mitigated with a reduction in the gate oxide thickness. These results help to elucidate the important role that contact geometry plays in the performance of carbon nanotube electronic devices.

Due to their unique structural and electrical properties, carbon nanotubes (CNTs) are promising candidates for next-generation electronic devices[1], and the CNT-metal contact plays a crucial role in their performance[2]. A variety of experimental[2-4] and theoretical[5-7] work has shown that the nature of the contact depends on the work function of the contact metal, and on the radius of the CNT. In particular, the Schottky barrier at the contact is inversely proportional to both the magnitude of the metal work function and the CNT radius, and the contact can become ohmic when these values become large enough[5,8]. The geometry of CNT electronic devices is also a factor in determining their performance. In field-effect transistors (FETs), the channel length and the gate oxide thickness are the traditional geometric parameters, and determine how well the charge in the channel is coupled to the potential on the gate. In addition to these parameters, the geometry of the CNT itself also plays an important role. Due to the unusual electrostatics and screening of quasi-1D structures, unique length scales and device performance can manifest themselves in CNT electronics[9].

In this work, we use numerical simulations to investigate the transport properties of CNT FETs. We find that competing electrostatic length scales result in unscreened charge on the CNT at the contact-channel interface. This leads to a potential barrier seen by charge carriers attempting to enter or leave the channel of the FET, which can significantly degrade transport through the transistor. This effect is largest for gate voltages near the turn-on point of the transistor, and for metals whose work function results in contacts near the crossover between being Schottky and ohmic. The effect can be mitigated by reducing the oxide thickness, which gives the gate greater control over the height of the contact-channel barrier.

The device to be simulated is a CNT FET as shown in Fig. 1, but the results apply to contacts to CNTs in general. The CNT sits on top of a dielectric a distance $t_{ox}$ above a gate electrode, and the ends of the CNT are embedded in the source and drain metals. For these simulations, we consider a (17,0) zig-zag nanotube with a diameter of 1.3 nm and a band gap of 0.54 eV. The dielectric is $SiO_2$, the channel length is 2 µm, the oxide thickness is 100 nm, the height of the source and drain contacts is 50 nm, and the CNT is separated from the oxide by 0.3 nm. There are 100 nm of vacuum between the top of the source/drain contacts and the upper edge of the simulation space. The length of the CNT-metal contact region is 100 nm, which is sufficiently long to ensure that the CNT energy bands are flat at the edges of simulation space (see Fig. 2), and that the CNT is in local equilibrium with the contact metal. The type of metal in the source and drain contacts is defined by the difference between its work function and that of the CNT, $\Delta\phi = \phi_{CNT} - \phi_{metal}$. The value of $\Delta\phi$ then determines the potential in the metallic contacts, assuming the reference potential is at the CNT mid-gap. In all cases, we assume a temperature of 300 K.

To determine the transport properties of the FET, we use a two-step procedure. The first step is to calculate the electric potential along the length of the CNT for a given gate voltage ($V_G$). This is accomplished with a self-consistent calculation of the charge and potential within the FET. The potential is obtained from the charge through a 3D solution of Poisson's equation,

$$\nabla \cdot (\varepsilon \nabla V) = -\rho, \qquad (1)$$

where $\rho$ is the charge density, $V$ is the electric potential, and $\varepsilon$ is the spatially-dependent dielectric constant. We treat the metals in the device as perfect electric conductors by imposing Dirichlet boundary conditions at the edges of the source, drain, and gate electrodes, and we assume Neumann boundary conditions at the left, right, and top edges of the simulation space shown in Fig. 1a. Periodic boundary conditions are applied along the *z*-axis, and we use a Fourier transform method to accelerate the solution of the 3D Poisson equation. The size of the supercell along this axis is chosen to avoid electrostatic interaction between neighboring CNTs[10]. Eq. (1) is discretized using the finite element method and is solved with a conjugate gradient algorithm, yielding a 3D potential profile, $V(x,y,z)$. The potential along the length of the CNT is then given by $V_{CNT}(x) = \frac{1}{2}[V(x, y_{top}, z_{top}) + V(x, y_{bot}, z_{bot})]$, where $(y_{top}, z_{top})$ are the coordinates on the top of the CNT in Fig. 1a, and $(y_{bot}, z_{bot})$ are the coordinates on the bottom of the CNT. We have found this gives the same results as taking an average potential over the entire circumference of the CNT. The charge density along the length of the CNT is calculated from the potential as

$$\rho_{CNT}(x) = \int_{-\infty}^{\infty} D(E + qV_{CNT}(x))f(E)dE, \qquad (2)$$

where $D(E)$ is the CNT density of states[11], $f(E)$ is the Fermi function, and $q$ is the electron charge. The 1D charge density on the CNT can then be mapped back to a 3D charge density, $\rho(x,y,z)$, by using a Gaussian distribution of the charge around the CNT radius. The width of the Gaussian distribution is 0.06 nm.

Once Eqs. (1) and (2) have been solved self-consistently, we describe the electronic structure of the CNT within a tight-binding representation[12], assuming one π-orbital per carbon atom and a coupling of $\gamma = 2.5$ eV between adjacent atoms[13,14]. We divide the zig-zag CNT into a series of layers, where each layer corresponds to a ring of carbon atoms. In this representation, the tight-binding Hamiltonian is given by $H_{ii} = -qV_{CNT}(x_i)$, $H_{2i,2i-1} = H_{2i-1,2i} = 2\gamma\cos(\pi J/M)$, and $H_{2i,2i+1} = H_{2i+1,2i} = \gamma$, where $x_i$ is the position of the $i^{th}$ carbon ring in the tight binding lattice, $J$ is the subband index, and $M$ is the number of atoms in each carbon ring. The zero-bias conductance through the device is then calculated as

$$G = \frac{4e^2}{h}\int T(E)\left[-\frac{df(E)}{dE}\right]dE, \qquad (3)$$

where $T(E)$ is the transmission through the device, and is calculated by applying the tight-binding Hamiltonian to the non-equilibrium Green's function (NEGF) formalism[15]. While we do not include additional scattering effects, the NEGF approach allows us to account for the elastic scattering that arises from variations in the potential along the length of the CNT.

Figure 2 shows the top of the CNT valence band for $V_G$ ranging from -15 V to + 5 V, with $\Delta\phi$ = -1 eV, corresponding to palladium contacts. The valence band energy is calculated from the self-consistent potential as $E_V(x) = -V_{CNT}(x) - E_g/2$, where $E_g$ is the CNT band gap. The Fermi level is indicated by the dotted line. In this figure, one can see that the contacts are ohmic and p-type. As shown in Fig. 2a, for positive values of $V_G$ the valence band in the channel is pulled far below the Fermi level, and the transistor is in the off state. For negative values of $V_G$, the valence band rises above the Fermi level, and holes can cross the channel from the source to the drain. However, at the interface between the contact and the channel, there is a "kink," or a dip, in the valence band, which acts as a barrier to the conduction of holes across the channel. This barrier can be seen more clearly in Fig. 2b, which shows a close-up view of the valence band at the contact-channel interface. Here, one can see that the height of this barrier is controlled by the gate, but much less strongly than the channel is controlled by the gate.

The existence of the contact barrier can be understood by considering the electrostatics of the FET geometry. Inside the contact, charge transfer between the metal and the CNT aligns their respective Fermi levels, and the result is a positive charge on the surface of the CNT (assuming $\Delta\phi < 0$, see Fig. 1). This charge is screened effectively by the surrounding metal of the contact, giving a relatively low potential. Just outside the contact, however, the charge on the CNT is no longer effectively screened by the contact metal. In addition, the contact metal shields the charge from the influence of the gate electrode. Therefore, in this intermediate region where neither the contact nor the gate can effectively screen the charge on the CNT, the potential increases, leading to the contact barrier. An examination of Fig. 2 reveals the two competing length scales involved in this phenomenon. The screening due to the contact is only effective within a distance on the order of the CNT radius ($\lambda_C \sim 0.66$ nm, see Fig. 2b), while the screening due to the gate is only effective at a distance greater than the oxide thickness ($\lambda_G \sim 100$ nm, see Fig. 2a).

In Fig. 3 we plot the conductance of a CNT FET as a function of the gate voltage, assuming $\Delta\phi = -1$ eV. The squares (black solid line) show the conductance including the contact barrier, while the triangles (red dashed line) show the conductance when the contact barrier has been artificially removed by bridging the potential in the contact with a constant value to the channel. The circles (blue dotted line) show the relative difference between these two curves, which indicates how much the contact barrier reduces the conductance of the CNT FET. For small gate voltages, charge transport is controlled by the potential in the middle of the channel, and the contact barrier has no effect. However, as the transistor starts to turn on, the transport becomes controlled by the electrostatic contact barrier. These two regimes of operation can be distinguished by a relatively sudden change in the slope of G vs. $V_G$, at $V_G$ = -1.3 V. The difference between the conductance with and without the contact barrier peaks at a value of 42%. This is the "flat-band" situation, where the potential in the middle of the channel equals that in the contacts. For any set of device parameters, the flat-band point is where the contact barrier has the largest effect on the conductance.

The work function of the contact metal also plays a role in the effect of the contact barrier on the transport. In Fig. 4a we plot the conductance as a function of $\Delta\phi$, with $V_G$ set to achieve the flat-band condition at each data point. This plot thus represents the upper bound on the effect of the contact barrier for the given CNT FET dimensions. One can see that this effect is largest for intermediate values of $\Delta\phi$, between -0.3 and -0.8 eV, where the relative difference between the two curves peaks at 50%. In this regime, the contacts go from being slightly ohmic to slightly Schottky, as seen in Fig. 4b, which shows the position of the valence band in the contacts for each value of $\Delta\phi$. For large values of $\Delta\phi$, the contact is strongly ohmic and the valence band lies far above the Fermi level, and thus the contact barrier has little effect on the charge transport. For small values of $\Delta\phi$, the contact is strongly Schottky and there is relatively little charge on the CNT in the contacts, which results in a small contact barrier.

As stated above, the contact barrier arises due to competition between the contacts and the gate, and their differing screening lengths. The simplest way to mitigate this effect, then, is to decrease $t_{ox}$, which will give the gate greater control over the height of the contact barrier. In Fig. 4a, with $t_{ox} = 100$ nm, the maximum reduction in the conductance due to the contact barrier was 50%. After decreasing the oxide thickness to 10 nm, we found that the maximum reduction in the conductance dropped to 25%. With additional decreases of $t_{ox}$, as well as the use of high-k dielectrics, this value would drop even further. But even in this case, it appears that the electrostatics of the contact play a significant role in the performance of CNT electronics.

This project is supported by the Laboratory Directed Research and Development program at Sandia National Laboratories, a multiprogram laboratory operated by Sandia Corporation, a Lockheed Martin Co., for the United States Department of Energy under Contract No. DEAC01-94-AL85000.

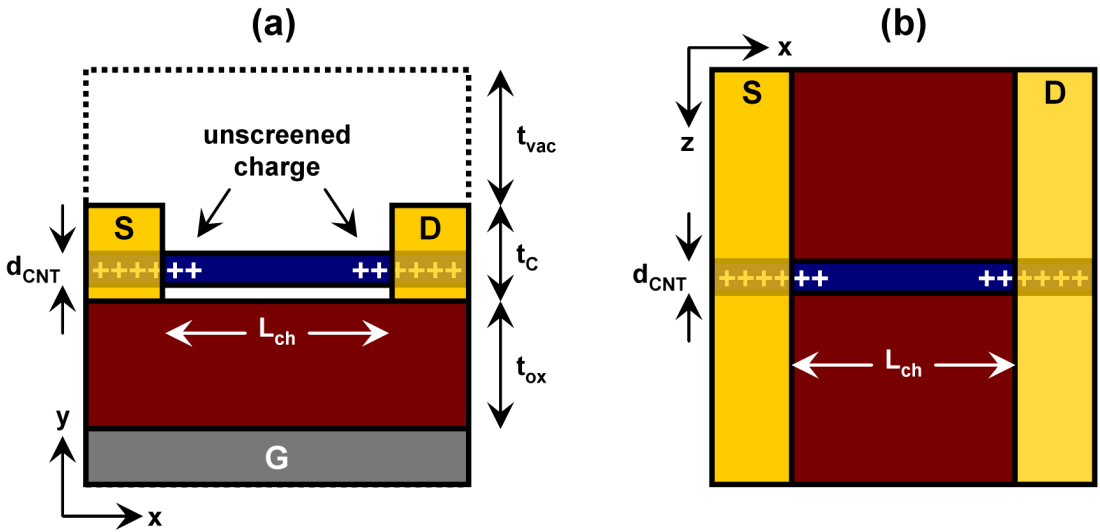

FIG. 1. Schematic of a carbon nanotube field-effect transistor, where part (a) shows the side view and part (b) shows the top view. The channel length ($L_{ch}$) is chosen to be 2 μm in this work, the oxide thickness ($t_{ox}$) is 100 nm, the contact metal thickness ($t_C$) is 50 nm, the nanotube diameter ($d_{CNT}$) is 1.3 nm, and the CNT is separated from the oxide by 0.3 nm. The distance from the top of the source/drain contacts to the upper edge of the simulation space ($t_{vac}$) is 100 nm.

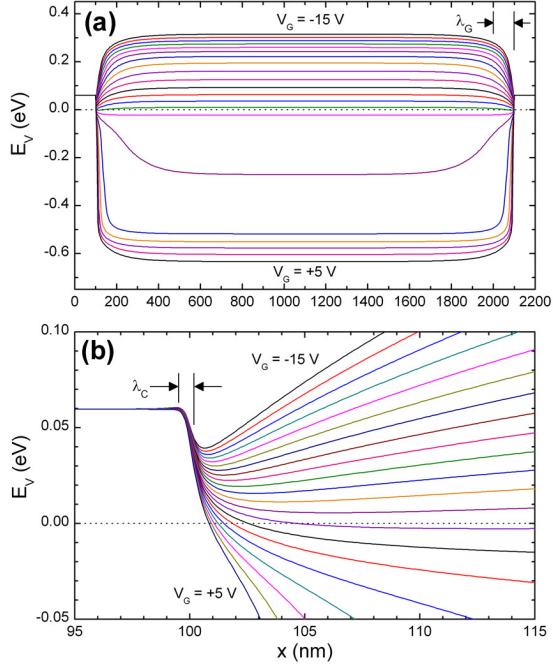

FIG. 2. Spatial dependence of the valence band of the CNT for various gate voltages, where $\Delta\phi = -1$ eV. Part (a) shows the entire device, and indicates the length scale, $\lambda_G$, over which the gate is screened by the contact metal. Part (b) shows a close-up of the contact-channel interface, which reveals the length scale, $\lambda_C$, over which the contact screens the charge on the CNT.

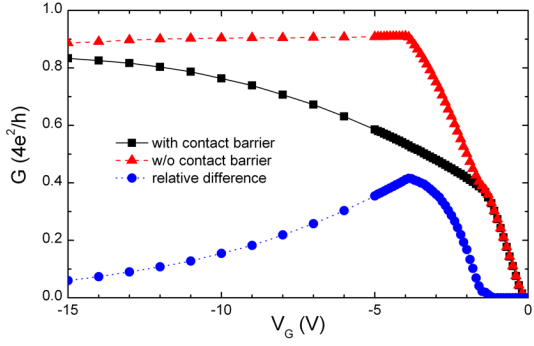

FIG. 3. Conductance vs. gate voltage of a CNT FET with (squares) and without (triangles) the contact barrier, and their relative difference (circles), where $\Delta\phi$ = -1 eV.

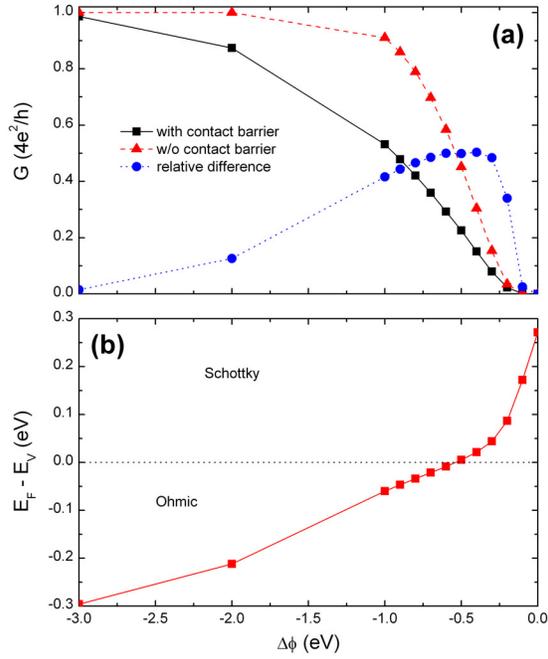

FIG. 4. Part (a) shows the conductance as a function of the CNT-metal work function difference (Δϕ) in the flat band condition. Part (b) shows the position of the valence band in the contacts, corresponding to each value of Δϕ.